\documentclass[]{jfm}

\usepackage{natbib}
\usepackage{graphicx}
\usepackage{amsmath}
\usepackage{amsfonts}

\ifCUPmtlplainloaded \else
  \checkfont{eurm10}
  \iffontfound
    \IfFileExists{upmath.sty}
      {\typeout{^^JFound AMS Euler Roman fonts on the system,
                   using the 'upmath' package.^^J}%
       \usepackage{upmath}}
      {\typeout{^^JFound AMS Euler Roman fonts on the system, but you
                   dont seem to have the}%
       \typeout{'upmath' package installed. JFM.cls can take advantage
                 of these fonts,^^Jif you use 'upmath' package.^^J}%
      }
  \else
  \fi
\fi

\ifCUPmtlplainloaded \else
  \checkfont{msam10}
  \iffontfound
    \IfFileExists{amssymb.sty}
      {\typeout{^^JFound AMS Symbol fonts on the system, using the
                'amssymb' package.^^J}%
       \usepackage{amssymb}%
       \let\le=\leqslant  
         
      }{}
  \fi
\fi

\ifCUPmtlplainloaded \else
  \IfFileExists{amsbsy.sty}
    {\typeout{^^JFound the 'amsbsy' package on the system, using it.^^J}%
     \usepackage{amsbsy}}
    {\providecommand\boldsymbol[1]{\mbox{\boldmath $##1$}}}
\fi

\renewcommand{\Re}{\mbox{\textit{Re}}}
\newcommand{\bu}{\boldsymbol{u}}
\newcommand{\bx}{\boldsymbol{x}}
\newcommand{\diam}{d_\mathrm{p}}
\newcommand{\xp}{\boldsymbol{X\!}_\mathrm{p}}
\newcommand{\vp}{\boldsymbol{V\!\!}_\mathrm{p}}
\newcommand{\omp}{\boldsymbol{\Omega}_\mathrm{p}}
\newcommand{\de}{\mathrm{d}}
\newcommand{\dd}[2]{\frac{\de{#1}}{\de{#2}}}
\newcommand{\DD}[2]{\frac{\mathrm{D}{#1}}{\mathrm{D}{#2}}}

\begin{document}

\title[Finite-size effects in the dynamics of neutrally buoyant
  particles in turbulent flow]{Finite-size effects in the dynamics of
  neutrally buoyant particles in turbulent flow}

\author[H.\ Homann and J.\ Bec]{H\ls O\ls L\ls G\ls E\ls R \ns H\ls
  O\ls M\ls A\ls N\ls N$^{1,2}$ \and J\ls\'{E}\ls R\ls\'{E}\ls M\ls
  I\ls E\ns B\ls E\ls C$^{2}$ }

\affiliation{$^1$ Theoretische Physik I, Ruhr-Universit\"at, 44780
  Bochum, Germany\\[\affilskip] $^2$ Universit\'e de Nice-Sophia
  Antipolis, CNRS, Observatoire de la C\^ote d'Azur,\break Laboratoire
  Cassiop\'ee, Bd.\ de l'Observatoire, 06300 Nice, France}

\date{\today}

\maketitle

\begin{abstract}
  The dynamics of neutrally buoyant particles transported by a
  turbulent flow is investigated for spherical particles with radii of
  the order of the Kolmogorov dissipative scale or larger. The
  pseudo-penalisation spectral method that has been proposed by
  \cite{PBC08} is adapted to integrate numerically the simultaneous
  dynamics of the particle and of the fluid. Such a method gives a
  unique handle on the limit of validity of point-particle
  approximations, which are generally used in applicative
  situations. Analytical predictions based on such models are compared
  to result of very well resolved direct numerical
  simulations. Evidence is obtained that Fax{\'{e}}n corrections give
  dominant finite-size corrections to velocity and acceleration
  fluctuations for particle diameters up to four times the Kolmogorov
  scale. The dynamics of particles with larger diameters is dominated
  by inertial-range physics, and is consistent with predictions
  obtained from dimensional analysis.
\end{abstract}

\section{Introduction}
\label{sec:intro}

A large number of natural and engineering situations involves the
transport of spherical finite-size particles by a fully developed
turbulent flow. This includes the formation of planets in the early
solar system, rain formation in clouds, the coexistence between
several species of plankton, and many industrial settings encountered
in chemistry and material processes.  An important feature of such
particles is that they do not follow exactly the fluid motion but have
inertia, a property that leads to the development of inhomogeneities
in their spatial distribution \citep[see][]{SE91,BFF01,BBC+07} or to
the enhancement of the rate at which they collide
\citep[see][]{FFS02,WMB06,ZSA06,BBC+09}. The modeling of such
particles generally assumes that their diameter $\diam$ is much
smaller than the smallest active length-scale of the flow, that is the
Kolmogorov scale $\eta$, so that they can be approximated by points
\citep[see][]{MR83,Gat83}. Modeling situations where
$\diam\gtrsim\eta$ relies on the use of various empirical laws
\citep[as reviewed, for instance, in][]{CGW78}. Generally such laws
are obtained by considering a particle suspended in a mean laminar
flow and interacting only through the turbulent wake that it creates,
but not with a fully developed turbulent environment maintained by an
external energy input.

Recent experimental developments have triggered a renewal of interest
in the understanding and quantification of finite-size effects in the
motion of particles in a turbulent flow
\citep{QBB+07,VCV+08,XB08,CVB+09}. These works addressed in particular
the problem of delimiting the domain of validity of the point-particle
models that are largely used in applicative fields, and to understand
which corrective terms give dominant corrections. Such questions
remain largely open because of the difficulty in constructing
analytically the fluid flow perturbed by the presence of the spherical
particle. In the following, we briefly review the equations that
govern the coupled dynamics of the flow and the particle. In an
incompressible fluid with kinematic viscosity $\nu$ and subject to an
external volumic forcing strain tensor $\mathbb{F}$, the velocity
field $\bu$ solves the Navier--Stokes equation
\begin{equation}
  \partial_t \bu + (\bu \cdot \nabla) \bu =
  -\frac{1}{\rho_\mathrm{f}}\,\nabla p + \nu\nabla^2 \bu +
  \nabla\cdot\mathbb{F}, \quad \nabla \cdot \bu = 0 ,
  \label{eq:ns}
\end{equation}
which is supplemented by a non-slip boundary condition
\begin{equation}
  \bu \left(\xp(t)+(\diam/2)\,\boldsymbol{n}, t\right) = \vp(t) +
  \frac{\diam}{2}\, \omp(t)\times\boldsymbol{n},\quad \forall
  \boldsymbol{n}:|\boldsymbol{n}| = 1
  \label{eq:boundary}
\end{equation}
at the surface $\partial \mathcal{B}$ of the spherical
particle. $\xp(t)$ denotes here the trajectory of the center of the
particle, $\vp(t)$ its translational velocity, and $\omp(t)$ its
rotation rate. The motion of the particle is determined by Newton's
second law
\begin{equation}
  m_\mathrm{p}\, \dd{\vp}{t} = (m_\mathrm{p}-m_\mathrm{f})\,
  \boldsymbol{g}+ \int_\mathcal{B} \nabla \cdot \mathbb{T} \,\,
  \de\mathcal{V} = (m_\mathrm{p}-m_\mathrm{f})\, \boldsymbol{g}
  +\int_{\partial \mathcal{B}}\mathbb{T} \cdot \de
  \boldsymbol{\mathcal{S}}
  \label{eq:newton}
\end{equation}
where $m_\mathrm{p} = (\pi/6) \rho_\mathrm{p}\, \diam^3 $ is the
particle mass (with $\rho_\mathrm{p}$ its mass density), $m_\mathrm{f}
= (\pi/6) \rho_\mathrm{f}\, \diam^3 $ the mass of the displaced fluid,
$\boldsymbol{g}$ is the acceleration of gravity
$\mathbb{T}=-p\,\mathbb{I}_3+({\mu}/{2})\,(\boldsymbol{\nabla}
  \bu+\boldsymbol{\nabla}\bu^\mathsf{T})+\rho_\mathrm{f}\,\mathbb{F}$
denotes the fluid stress tensor, $\mathbb{I}_3$ is the identity, and
$\mu = \rho_\mathrm{f}\nu$ the dynamic viscosity. In addition, the
sphere rotation rate $\omp$ changes according to the conservation of
angular momentum
\begin{equation}
  \mathcal{I}\, \dd{\omp}{t} = \frac{m_\mathrm{p}}{10} \,\diam^2\,
  \dd{\omp}{t} = \int_{\partial \mathcal{B}} \boldsymbol{n}
  \times(\mathbb{T} \cdot \de\boldsymbol{S}),
  \label{eq:momentum}
\end{equation} 
where $\boldsymbol{n}$ denotes the outward pointing unit-vector normal
to the surface and where we have assumed that the mass moment of
inertia tensor $\mathcal{I}$ is that of a uniform solid
sphere. Solving the system (\ref{eq:ns})-(\ref{eq:momentum}) is a
difficult task as it involves a nonlinear partial differential
equation for the fluid, which is coupled to a moving boundary
condition on the sphere. Analytical treatments of such a complex
dynamics has only been done when neglecting nonlinearities in the flow
motion at the scale of the particle, so that (\ref{eq:ns}) reduces to
the Stokes equation (this leads to the usual point-particle models).

This study focuses on neutrally buoyant particles, \textit{i.e.}\/
$\rho_\mathrm{p} = \rho_\mathrm{f}$. This case is of interest for
applications to problems of plankton dynamics in the ocean or of some
types of ice crystals in clouds. The goal is here to give a complete
description of the dynamical properties of particles with sizes of the
order of the Kolmogorov dissipative scale $\eta$. The paper is
organised as follows.  In \S\ref{sec:point-particle}, we consider the
model given by the point particle approximation. We show that in the
case of neutrally buoyant particles, first-order finite-size effects
are not due to particle inertia but purely stem from Fax{\'{e}}n
corrections. They intervene in the particle dynamics as
$(\diam/\lambda)^2$, where $\lambda$ designates the Taylor
micro-scale. These results are validated numerically in
\S\ref{sec:numerics} thanks to the use of a new dynamical
pseudo-penalisation technique that has the advantage of allowing one
to use a spectral code in order to integrate the Navier--Stokes
equation with the proper boundary conditions. We show that, both for
velocity and acceleration statistics, finite-size effects become
noticeable for $\diam \simeq 3\,\eta$ and that first-order Fax\'en
corrections are relevant up to $\diam = 4\,\eta$. For $\diam \gtrsim
4\,\eta$, the particle dynamics is dominated by inertial-range
physics. We also present results on acceleration time correlation that
confirm this fact. Finally, \S\ref{sec:conclusion} is dedicated to
concluding remarks and prospectives.

\section{Point-particle approximation}
\label{sec:point-particle}
The derivation of point-particle models relies on the assumption that
the perturbation of the surrounding flow by particles is well
described by the Stokes equation \citep[see][]{Gat83,MR83,AHP88}. This
assumption clearly requires that the particle Reynolds number defined
with the velocity difference between the fluid and the particle is
very small. The motion of the neutrally buoyant particle is then given
by
\begin{equation}
  \dd{\vp}{t} = \boldsymbol{A}^\mathcal{V}(t) -\frac{12\nu}{\diam^2}
  \left( \vp - \boldsymbol{U}^\mathcal{S}(t) \right) +
  \frac{6}{\diam}\sqrt{\frac{\nu}{\pi}}\int_{-\infty}^t \left(
  \boldsymbol{A}^\mathcal{V}(s) - \dd{\vp}{s}
  \right)\frac{\de s}{\sqrt{t-s}},
  \label{eq:fullpoint}
\end{equation}
where $\boldsymbol{U}^\mathcal{S}$ and $\boldsymbol{A}^\mathcal{V}$
account for Fax\'en corrections. They are averages of the fluid
velocity over the surface and of the fluid acceleration over the
volume of the particle, respectively:
\begin{equation}
  \boldsymbol{U}^\mathcal{S}(t) = \frac{2}{\pi\, \diam^2}
  \int_{\partial\mathcal{B}} \bu(\bx,t) \,\de\mathcal{S} \quad
  \mbox{and}\quad\boldsymbol{A}^\mathcal{V}(t) = \frac{6}{\pi\,
    \diam^3} \int_\mathcal{B} \DD{\bu}{t}(\bx,t)
  \,\de\mathcal{V},
\end{equation}
where $\mathrm{D}/\mathrm{D}t = \partial_t +\bu\cdot\nabla$ denotes
the material derivative along fluid tracer trajectories.  The various
forcing terms in (\ref{eq:fullpoint}) are, in order of appearance, the
combination of the inertia force exerted by the undisturbed flow and
the added mass, the Stokes viscous drag, and the Basset--Boussinesq
history force.  In the limit when the particle size is much smaller
than the Taylor micro-scale $\lambda$, a Taylor expansion of the fluid
velocity in the vicinity of the particle center leads to
\begin{equation}
  \begin{cases}
    \boldsymbol{U}^\mathcal{S}(t) = \bu(\xp,t) +
    \frac{1}{40}\,\diam^2\, \nabla^2 \bu(\xp,t) +
    \mathcal{O}[(\diam/\lambda)^4], \\ \boldsymbol{A}^\mathcal{V}(t) =
    (\mathrm{D}\bu/\mathrm{D}t)(\xp,t) + \frac{1}{24}\,\diam^2\,
    (\mathrm{D}\nabla^2 \bu/\mathrm{D}t)(\xp,t) +
    \mathcal{O}[(\diam/\lambda)^4].
    \end{cases}
  \label{eq:expandfaxen}
\end{equation}
In the limit of particle diameters much smaller than the Kolmogorov
scale $\eta$, the Basset--Boussinesq history force gives a
contribution much smaller than the viscous drag and can thus be
neglected to leading order.  Hence, finite-size neutrally buoyant
particles obey asymptotically the minimal model equation
\begin{equation}
\label{eq:point_model}
  \dd{\vp}{t} = \DD{\bu}{t}(\xp,t) -\frac{12\nu}{\diam^2}
  \left[\vp-\bu(\xp,t)\right].
\end{equation}
Note that in this model, the particle size enters only the coefficient
of the drag force. However, as it is now shown, such an effect is
actually not sufficient to account for leading-order corrections due
to the particle finite size. Indeed, following \cite{BCP+00} and
introducing the velocity difference between the particle and the fluid
$\boldsymbol{W}(t) = \vp(t) - \bu(\xp(t),t)$, one can easily check
that
\begin{equation}
  \dd{\boldsymbol{W}}{t} = -\boldsymbol{W}\cdot\nabla\bu(\xp(t),t)
  -\frac{12\nu}{\diam^2} \boldsymbol{W}.
\end{equation}
This implies that $\boldsymbol{W}(t) = \exp(-12\nu\,t/\diam^2)\,
\mathcal{T}\!\!\exp [-\int_0^t\nabla\bu(\xp(s),s)\,\de s]
\,\boldsymbol{W}(0)$, where $\mathcal{T}\!\!\exp$ denotes the
time-ordered exponential. Hence the amplitude of the velocity
difference grows exponentially at large time, \textit{i.e.}\/
$|\boldsymbol{W}(t)| \simeq
|\boldsymbol{W}(0)|\,\exp[-(12\nu/\diam^2+\lambda_3)\,t]$, where
$\lambda_3$ is the smallest Lyapunov exponent associated to
$\mathcal{T}\!\!\exp [-\int_0^t\nabla\bu(\xp(s),s)\,\de s]$. Because
of the fluid flow incompressibility implying a vanishing sum of the
Lyapunov exponents, $\lambda_3\le 0$. However, when $\diam$ is
sufficiently small, \textit{i.e.}\/ when $\diam \le
\sqrt{12\nu/|\lambda_3|}$, the exponential growth rate of
$|\boldsymbol{W}|$ is negative and the particle velocity relaxes to
that of the fluid. Estimating the value of $\lambda_3$ requires in
principle to evaluate the Lyapunov exponents associated to the fluid
flow strain along particle trajectories. However, because of the
exponential relaxation of the particle velocities to that of the
fluid, these exponents are exactly those computed along tracer
trajectories. The latters have been evaluated in direct numerical
simulations \citep[see, \textit{e.g.},][]{BBB+06} and, once normalised
by the inverse of the Kolmogorov eddy turnover time $\tau_\eta$,
depend weakly upon the Reynolds number of the flow: for $\Re_\lambda$
varying from 65 to 185, one observes $\tau_\eta \lambda_3 =
0.190\pm2\%$. To summarise, this shows that the minimal model
(\ref{eq:point_model}) exactly sticks to tracer dynamics for sizes
smaller than a fixed threshold, \textit{i.e.}\/ for
\begin{equation}
  \Phi = \frac{\diam}{\eta} \le \Phi^\star =
  \sqrt{\frac{12}{\tau_{\eta}\lambda_3}}\approx 8.
  \label{eq:phicrit}
\end{equation}
Hence, for $\Phi \le \Phi^\star$ finite-size effects are not related
to inertia but can only stem from terms that were neglected in this
model. This observation explain why neutrally buoyant particles whose
dynamics is approximated by the point-particle model
(\ref{eq:point_model}) do not display any clustering properties, as
observed by~\cite{CKL+08}.

Let us now estimate the contribution from terms that were neglected,
namely from the Fax\'en corrections and the Basset-Boussinesq history
term. For this, we follow the approach of \cite{M87} developed for
small Stokes numbers and write the following perturbative Ansatz
$\vp(t) = \bu(\xp(t),t) + (\diam/\lambda)^\alpha \boldsymbol{f}(t) +
\mathrm{o}[(\diam/\lambda)^\alpha]$, where the order $\alpha$ and the
function $\boldsymbol{f}$ are to be determined. Inserting this form in
(\ref{eq:fullpoint}) and (\ref{eq:expandfaxen}), one obtains that the
first-order terms originate from Fax\'en corrections to the viscous
drag, so that $\alpha=2$ and
\begin{equation}
  \vp(t) = \bu(\xp(t),t) + \frac{\diam^2}{40}\,\nabla^2\bu(\xp(t),t)
  +\mathcal{O}[(\diam/\lambda)^4].
  \label{eq:asymptfaxen}
\end{equation}
Note that the synthetic velocity field defined above is
divergence-free, implying that no effect of particle preferential
concentration can be detected with first-order corrections. This
asymptotic form (\ref{eq:asymptfaxen}) implies that the particle
velocity variance satisfies
\begin{equation}
  \left\langle |\vp|^2 \right\rangle - \left\langle |\bu|^2
  \right\rangle \simeq \frac{\diam^2}{20} \left\langle\bu\cdot
  \nabla^2\bu \right\rangle = - \frac{\diam^2}{20}
  \frac{\varepsilon}{\nu} = - \frac{\left\langle |\bu|^2
    \right\rangle}{20} \left(\frac{\diam}{\lambda}\right)^2,
  \label{eq:varvelofaxen}
\end{equation}
where $\varepsilon$ is the mean turbulent rate of kinetic energy
dissipation.  As for the variance of particle acceleration, one
obtains from the time derivative of (\ref{eq:asymptfaxen})
\begin{equation}
  \left\langle \left|\dd{\vp}{t}\right|^2 \right\rangle - \left\langle
  \left|\DD{\bu}{t}\right|^2 \right\rangle \simeq - \frac{\diam^2}{20}
  \left\langle \left\|\DD{\boldsymbol{\nabla}\bu}{t}\right\|^2
  \right\rangle = - \frac{\diam^2}{20} \left\langle
  (\nabla^2p)^2+\nu^2
  \left\|\nabla^2\boldsymbol{\nabla}\bu\right\|^2\right\rangle,
  \label{eq:varaccelfaxen}
\end{equation}
where $\|\cdot\|$ denotes the tensorial Frobenius norm,
\textit{i.e.}\/ $\|\mathbb{M}\|^2 = \mathrm{trace}\,(
\mathbb{M}^\mathsf{T}\, \mathbb{M})$. Hence we expect at small
diameters that finite-size effects materialise as a falloff of the two
above-mentioned dynamical properties of particles that behave
quadratically as a function of the particle diameter with a
coefficient given by Eulerian averaged quantities.

\section{Numerical results using a pseudo-penalisation method}
\label{sec:numerics}

In order to assess numerically the effect of the particles' finite
size on their dynamics, a pseudo-spectral code has been adapted to
non-trivial geometries by using a pseudo-penalisation strategy. The
purely spectral part of the parallel code LaTu, which was already used
to investigate Lagrangian turbulence by~\cite{HGB+07}, solves
accurately the incompressible Navier-Stokes equations. This method
treats the evolution of the fluid velocity field in Fourier space and
computes convolutions arising from the non-linear terms in physical
space. The Fourier transformations are performed by the
P3DFFT-library.\footnote{Parallel 3D Fast Fourier Transforms (P3DFFT),
  http://www.sdsc.edu/us/resources/p3dfft} The domain is a
triple-periodic cube. This method allows for high accuracy, precise
control of the physical parameters and numerical efficiency. In order
to maintain a statistically steady state we force the flow by
prescribing the energy content of the Fourier vectors with moduli 1
and 2.  The energy content of each of these two shells is kept
constant while the individual amplitudes and phases are evolved
piecewise linearly in time between several random configurations
separated by a time $10\,T_L$. The advantages of such a forcing are
two-fold: it allows one to achieve a statistically isotropic
large-scale flow and limits the fluctuations to only approximatively
$10\%$ of the mean. The turbulent characteristics of the flow
generated in this way are summarised in Tab.~\ref{table}.
\begin{table}
  \begin{center}
    \begin{tabular}{llllllllllllll}
      $\Re_{\lambda}$&$u_\mathrm{rms}$& $\varepsilon$ &$\nu$ &
      $\delta x$ & $\delta t$ & $\eta$ &$\tau_\eta$&$L$&$T_L$& $N^3$
      \\ $32$ &$0.17$ &$4.5\cdot 10^{-3}$&$3 \cdot 10^{-3}$&$1.23\cdot
      10^{-2}$& $8\cdot10^{-3}$ & $5 \cdot 10^{-2}$& 0.8 &1.2& 6.5
      &$512^3$
    \end{tabular}
    \caption{\label{table} Parameters of the numerical simulations.
      $\Re_\lambda = \sqrt{15u_\mathrm{rms}L/ \nu}$: Taylor-Reynolds
      number, $u_\mathrm{rms}$: root-mean-square velocity,
      $\varepsilon$: mean kinetic energy dissipation rate,
      $\nu$: kinematic viscosity, $\delta x$: grid-spacing, $\delta
      t$: time step, $\eta =(\nu^3/\varepsilon)^{1/4}$:
      Kolmogorov dissipation length scale, $\tau_\eta =
      (\nu/\varepsilon)^{1/2}$: Kolmogorov time scale, $L =
      (2/3E)^{3/2}/\varepsilon$: integral scale, $T_L =
      L/u_\mathrm{rms}$: large-eddy turnover time, $N^3$: number of
      collocation points.}
  \end{center}
\end{table}

\begin{figure}
  \begin{center}
    \begin{minipage}{0.43\textwidth}
      \includegraphics[width=\textwidth]{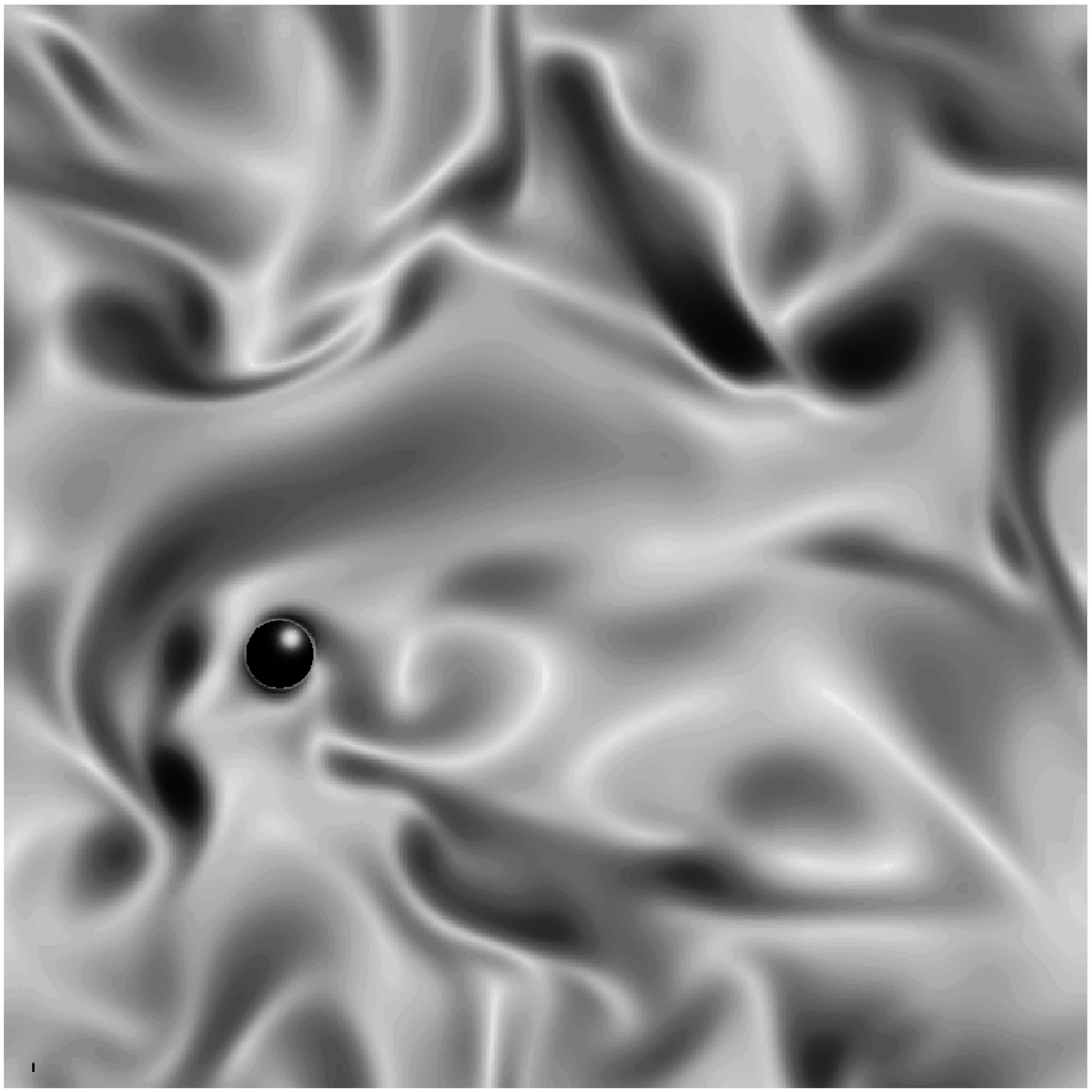}\\[6pt] \strut
    \end{minipage}
    \begin{minipage}{0.55\textwidth}
      \includegraphics[width=\textwidth]{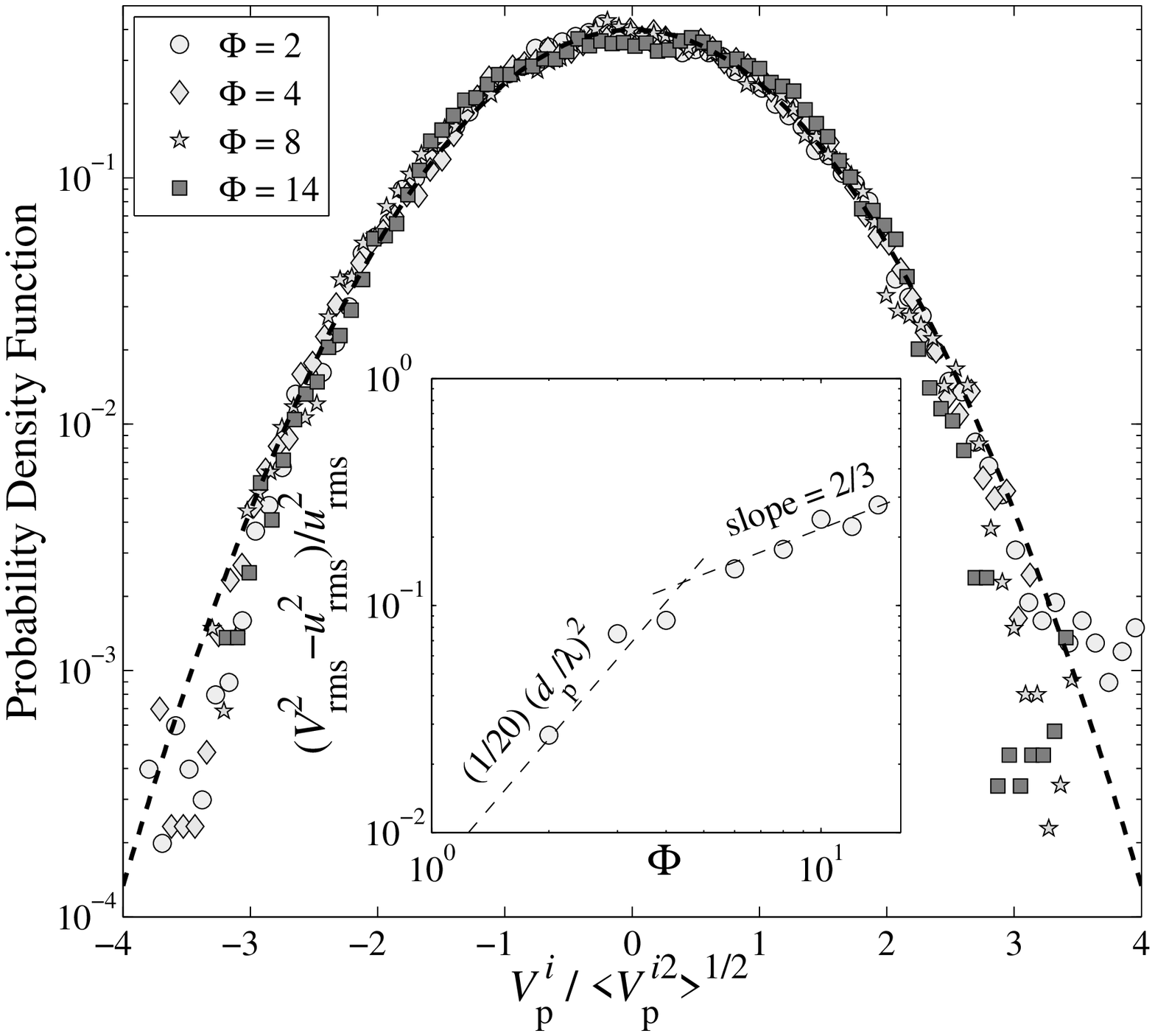}
      \end{minipage}
  \end{center}
  \caption{\emph{Left:} modulus of the vorticity in a slice of the
    domain that is passing through the center of the embedded particle
    (dark = high vorticity, light = low vorticity); the particle
    diameter is here $\diam = 8\eta$. \emph{Right:} normalised
    probability density function of the particle velocity for various
    particles sizes, as labeled; the bold dash line corresponds to a
    Gaussian distribution. Inset: deviation of the particle velocity
    variance $V_\mathrm{rms}^2 = \langle |\vp|^2\rangle$ from the
    fluid value, as a function of the non-dimensionalised particle
    diameter $\Phi = \diam/\eta$; the two dashed line represent the
    deviation (\ref{eq:varvelofaxen}) from the fluid root-mean square
    velocity that is expected to stem from Fax\'en corrections and a
    behaviour $\propto \diam^{2/3}$, respectively .}\label{fig:cut}
\end{figure}

In order to impose the no-slip boundary condition on the surface of
the spherical particle, we follow \cite{PBC08} and use a
pseudo-penalisation method, which consists in imposing a strong drag
to the fluid velocity at the particle location, so that it relaxes
quickly to the particle solid motion. The hydrodynamical forces acting
on the particle are computed by a Riemann approximation of the
integrals appearing in (\ref{eq:newton}) and (\ref{eq:momentum}) on a
homogeneous grid of discrete points located on the surface of the
sphere. The value of pressure at these points is computed by tri-cubic
interpolation. The surface integral of the fluid velocity gradient is
computed from evaluating the average velocity on spherical shells
surrounding the particle. At the moment, the simulations are limited
to a single particle in order to prevent the individual dynamical
properties from being contaminated by particle-particle hydrodynamical
interactions. Notice however that the code is very well adapted to
situations involving several particles. Because only a single isolated
particle is considered in the flow, the statistical convergence of
particle-related quantities requires to perform averages over very
long times. Each simulation for a single value of the particle
diameter required to integrate the flow over more than three hundreds
large-eddy turnover times. Eight different particle diameters $\diam$
are considered within the range $2\,\eta$ to $14\,\eta$. As the
pseudo-penalisation technique requires several grid-points inside the
object in order to correctly impose the boundary conditions, the
Kolmogorov dissipative scale $\eta$ is resolved with four grid
points. This requires the use of double floating point precision.
Because of the large spatial resolution and the long time integration
which are required, the simulations are very computationally
demanding: this work took approximatively four millions of single
processor CPU hours.  Figure~\ref{fig:cut} (Left) shows the typical
vorticity field in a cut-plane passing through the center of the
particle. Note that the signature of a turbulent wake is visible on
the right-hand side of the particle.

To benchmark these simulations, a run with the same parameters as
those shown in Tab.~\ref{table} but without any finite-size particle
has been performed. In this simulation, we have integrated the motion
of passive point particles with a dynamics
obeying~(\ref{eq:point_model}) and of passive tracers. It is
worthwhile mentioning here that turbulent fluid statistical quantities
are observed not to depend on the presence of a particle, up to the
statistical fluctuations due to finite time averages. The
particle-free simulation serves thus as a reference to estimate tracer
statistics and Eulerian averages.

Figure~\ref{fig:cut} (Right) shows the probability density function
(PDF) of the particle velocity components for various particle
diameters. Once normalised by their standard deviations, these PDFs
almost collapse on top of each other and deviate very weakly from
a Gaussian distribution. The measured variance of the particle
velocity that is represented in the inset, decreases as a function of
the particle size. For small diameters, \textit{i.e.}\/ for $\Phi =
\diam/\eta\lesssim 4$, the behaviour of the particle velocity variance
is very well described by the prediction (\ref{eq:varvelofaxen})
obtained from Fax\'en corrections. For $\Phi\gtrsim 4$, the deviation
from the fluid velocity variance behaves as $\Phi^{2/3}$. This
power-law is dimensionally compatible with Kolmogorov 1941 scaling and
indicates that particles with such diameters respond to the
inertial-range physics of turbulence. Noticeably, in the
low-Reynolds-number flow that we are considering here, there is no
inertial range in the sense usually defined through velocity scaling
properties. Hence it seems that particle dynamical properties are much
more amenable to dimensional estimates than fluid turbulent
quantities.

\begin{figure}
  \centerline{\includegraphics[width=0.5\textwidth]
    {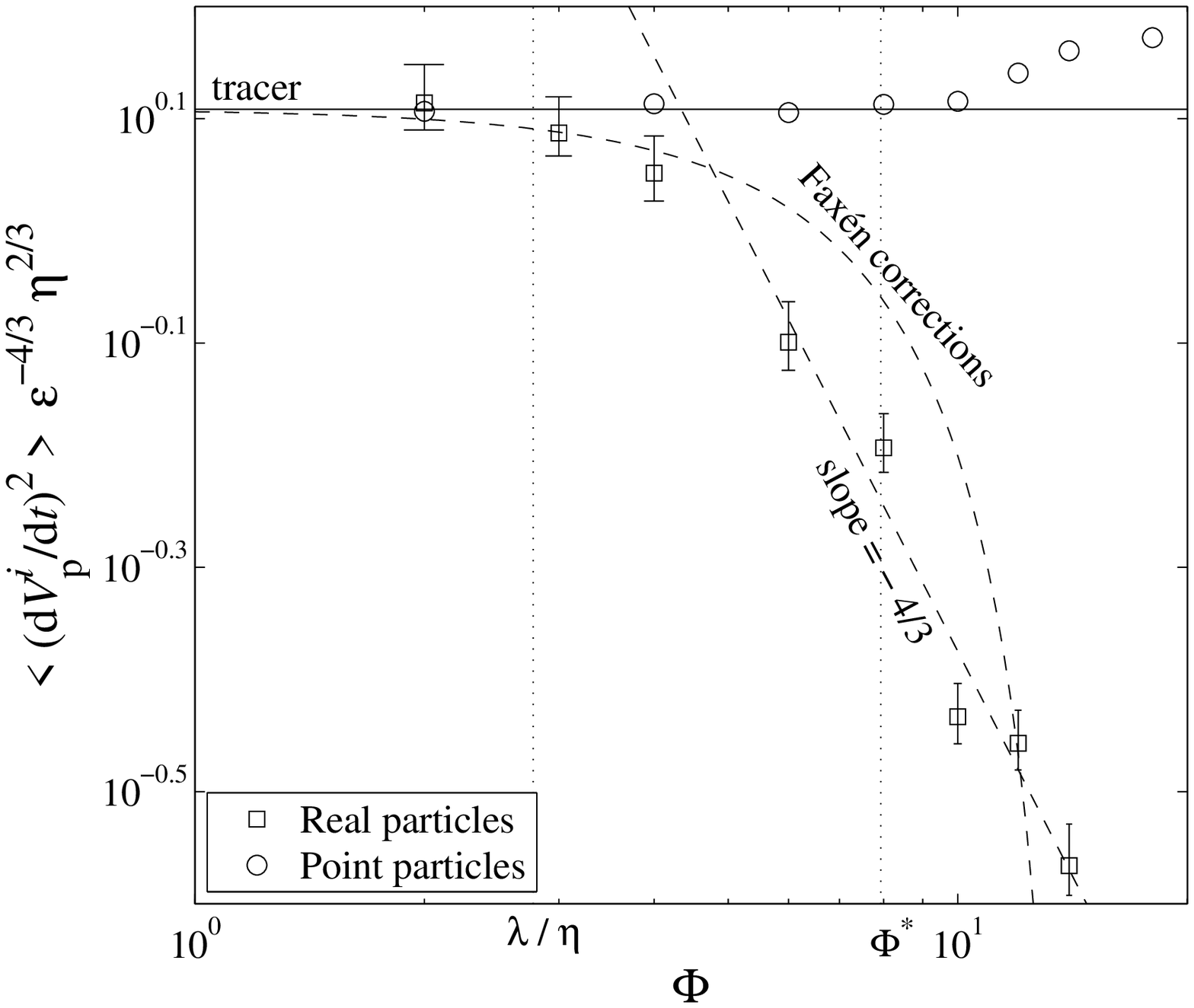}
    \includegraphics[width=0.5\textwidth]{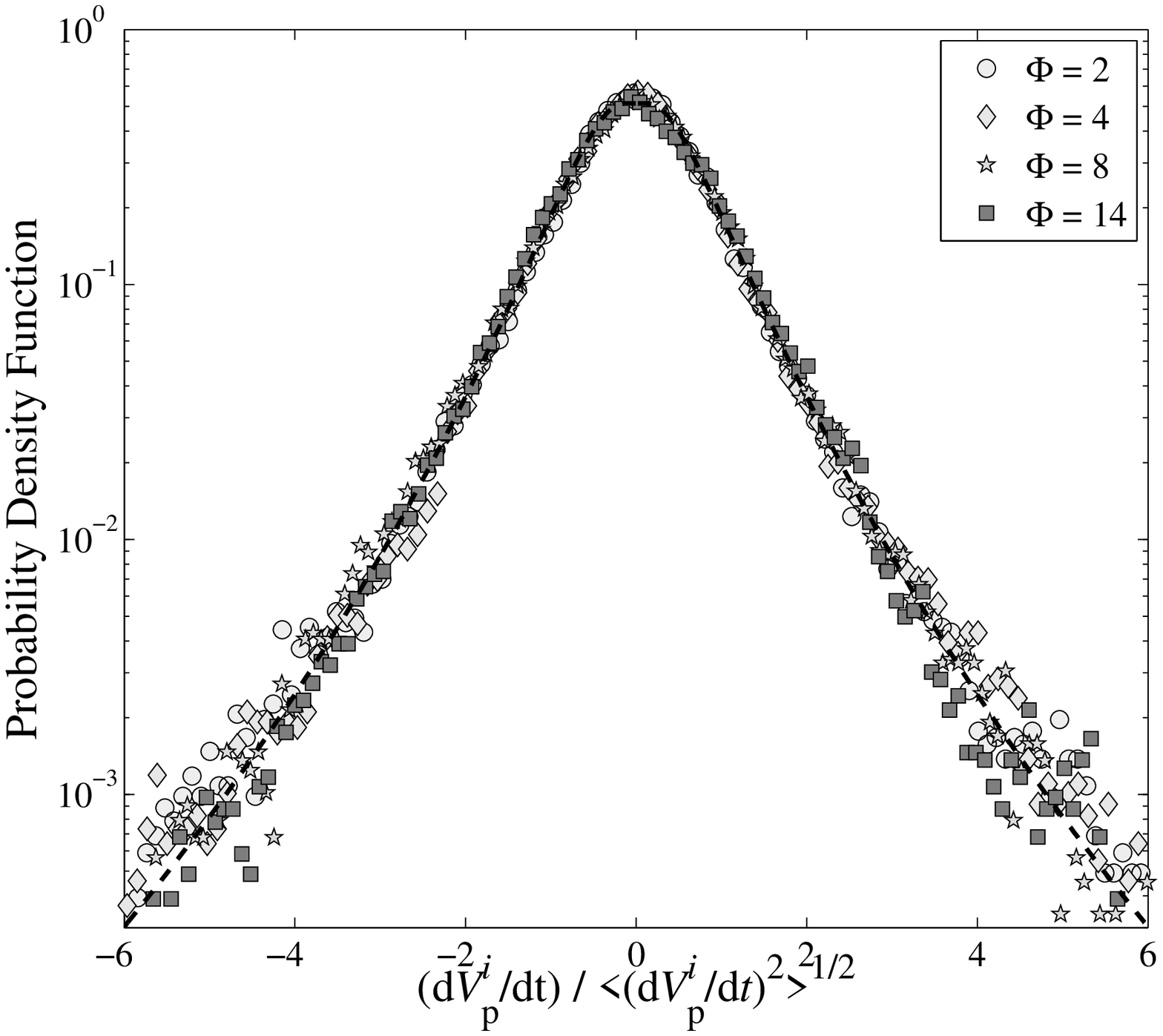}}
  \caption{\emph{Left:} acceleration variance of real spheres and of
    point particles as a function of the non-dimensionalised particle
    diameter $\Phi = \diam/\eta$; error bars correspond to an
    estimation of the standard deviations.  The two dotted vertical
    line indicate the Taylor micro-scale $\lambda$ and the critical
    value (\ref{eq:phicrit}) above which point particles deviate from
    fluid tracers.  The dashed curve corresponds to the
    behaviour~(\ref{eq:varaccelfaxen}) predicted from Fax\'en
    corrections.  \emph{Right:} normalised probability density
    function of the component-wise particle acceleration for various
    particle diameter, as labeled. The bold dashed line on which data
    almost collapse corresponds to the log-normal fit
    (\ref{eq:lognorm}) of experimental data proposed
    by~\cite{QBB+07}.}\label{fig:rms_comp}
\end{figure}
We next turn to particle acceleration statistics.
Figure~\ref{fig:rms_comp} (Left) represents the componentwise
normalised variance of the particle acceleration $a_0 = \langle
(\de{V_\mathrm{p}^i}/{\de t})^2\rangle \epsilon^{-4/3} \eta^{2/3}$ as
a function of the non-dimensionalised diameter $\Phi = \diam/\eta$,
both for the real particles as well as for the minimal point
model~(\ref{eq:point_model}). For real spherical particles, one
distinguishes, as in the case of velocity variance, between two
behaviours. When $\Phi = \diam/\eta \lesssim 4$, finite-size effects
in the acceleration variance are very well captured by Fax\'en
corrections and are very close to the prediction
(\ref{eq:varaccelfaxen}). Note that, thanks to their isotropic form,
the sub-leading terms appearing in (\ref{eq:varaccelfaxen}) could be
evaluated here through the pressure and velocity spectra. When
$\Phi\gtrsim 4$, an inertial-range behaviour with $a_0\propto
\diam^{-4/3}$ is attained. As argued by~\cite{QBB+07}, the variance of
finite-size particle acceleration is related to that of the fluid
pressure integrated over a sphere of diameter $\diam$. The power
$-4/3$ that is observed here differs from the value $-2/3$, which was
measured by Qureshi \textit{et al.}  However, as already stressed for
instance in \cite{BBC+07}, pressure scaling in low Reynolds number
flows is often dominated by sweeping, leading to a behaviour of the
pressure increments $|p(x+\ell)-p(x)|\sim\ell^{1/3}$. While no scaling
of pressure can be detected in the present simulations, the robustly
observed $-4/3$ law can be explained with such a sweeping-dominated
pressure spectrum. Another observation is that numerics confirm the
presence of the threshold (\ref{eq:phicrit}) predicted in the previous
section for the minimal point particle model: indeed the numerical
integration of point particles obeying (\ref{eq:point_model}) shows
that when $\Phi<8$, the acceleration variance of the latters is
undistinguishable from that of tracers. For $\Phi>8$, the point
particle model gives an enhancement of acceleration, which is
incompatible with measurements done with real particles at the
Reynolds number considered here. This stresses the irrelevance of such
a model in the case of neutrally buoyant particles.  Note finally that
in the case of tracers, the constant $a_0$ is known to show a Reynolds
number dependence $a_0 \propto R_\lambda^{1/2}$ \citep[see,
  \textit{e.g.},][]{VLC+02}. The measured value of approximatively
$1.3$ is in good agreement with the value that was measured
experimentally by \cite{QBB+07}.

Figure~\ref{fig:rms_comp} (Right) represents the PDF of acceleration
components normalised to unity variances for various values of the
particle diameter. As already stressed in \cite{QBB+07}, the
dependence upon $\Phi=\diam/\eta$ is very weak. Data can be fitted by
the function
\begin{equation}
  p(a) = \left[\exp\left(3s^2/2\right)/\left(4\sqrt{3}\right) \right]
  \left\{ 1 - \mathrm{erf}\left[\left(\ln|x/\sqrt{3}| +
    2s^2\right)/\left(\sqrt{2}s\right)\right) \right\},
\label{eq:lognorm}
\end{equation}
which was proposed by \cite{MCB04} for the acceleration PDF of fluid
tracers. Numerical results almost collapse to such a distribution with
a value of the parameter $s=0.62$, as observed by \cite{QBB+07}.

\begin{figure}
  \centerline{\includegraphics[width=0.9\textwidth]{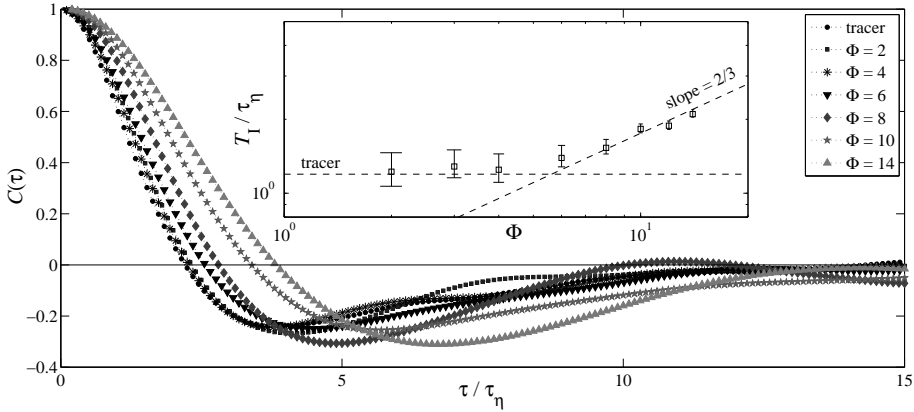}}
  \caption{Component-wise acceleration time correlation $C(\tau)$
    defined in (\ref{eq:defcorrel}) for various particle sizes, as
    labeled. Inset: integral correlation time $T_\mathrm{I}$ defined
    from (\ref{eq:defTI}) as a function of the non-dimensionalised
    particle diameter $\Phi =
    \diam/\eta$.} \label{fig:correlation_A_comp}
\end{figure}

Other measurements relate to two-time statistical properties of
particles. Figure~\ref{fig:correlation_A_comp} represents the
acceleration time correlation 
\begin{equation}
  C(\tau) \equiv \left \langle \dd{V_p^i}{t}(t+\tau) \dd{V_p^i}{t}
  (t) \right\rangle / \left\langle
  \left(\dd{V_p^i}{t}\right)^2\right\rangle
  \label{eq:defcorrel}
\end{equation}
as a function of the time lag $\tau$ for various values of the
particle diameter. The numerical measurements reported here are in
good agreement with the results of \cite{CVB+09}.  Surprisingly one
observes that $C(\tau)$ deviates only very weakly from the tracer
acceleration temporal correlation for diameters less than $4\eta$,
that is when Fax\'en corrections are expected to be of relevance to
capture first-order finite-size effects. This behaviour is even
clearer when looking at the diameter dependence of the correlation
time for particle acceleration. For this we follow \cite{CVB+09} and
introduce the integral time
\begin{equation}
  T_\mathrm{I} \equiv \int_0^{T_0} C(\tau)\,\de\tau,
  \label{eq:defTI}
\end{equation}
where $T_0$ is the first zero-crossing time. The inset of
Fig.~\ref{fig:correlation_A_comp} represents $T_\mathrm{I}/\tau_\eta$
as a function of $\Phi = \diam/\eta$. When $\Phi\lesssim4$, this
integral correlation time is, up to numerical errors,
undistinguishable from the value obtained for tracers. When
$\Phi\gtrsim4$, the correlation time increases much faster as a
function of $\Phi$ and follows approximatively the power-law behaviour
$T_\mathrm{I}\sim\Phi^{2/3}$. This indicates again that turbulent
inertial physics is pulling strings at such values of the particle
diameter, and that the relevant time scale is then given by the eddy
turnover time $\sim\varepsilon^{-1/3}\diam^{2/3}$ associated to the
particle size.

\section{Concluding remarks}
\label{sec:conclusion}

In this paper it is shown that first-order finite-size corrections to
the dynamics of neutrally buoyant particles are due to Fax\'{e}n terms
and not to particle inertia. This leads to several predictions on
turbulent velocity and acceleration second-order statistics. Using a
pseudo-penalisation spectral method, these predictions have been
confirmed numerically for particles with diameters $\diam$ up to
$4\,\eta$. Higher-order statistics of velocity and acceleration seem
much less sensitive to the finiteness of the particle sizes, since
numerical observation suggest that, once normalised to unit variance,
their PDFs collapse on top of each other for various values of the
particle diameter.

The irrelevance of particle inertia with respect to Fax\'{e}n terms at
small particle diameters has noticeable consequences. First, it
implies that the particle dynamics is very well approximated by the
advection by a synthetic flow, which is incompressible to both the
leading and the first orders. This means that the effect of
preferential concentration onto the dynamics of neutrally buoyant
particles is very weak. Secondly, such an observation clearly
questions the relevance of inertial-particle models for density
contrasts between the particle and the fluid that are close to one.  A
third consequence is related to the fact that corrective terms apply
when the particle diameter is much smaller than the Taylor micro-scale
$\lambda$ rather than the Kolmogorov dissipative scale $\eta$. This
fact might partly explain the difficulties in matching experiments and
numerical model including Fax\'en corrections \citep[see,
  \textit{e.g.},][]{CVB+09}.

For particle diameters larger than $\approx\!4\,\eta$, inertial-range
physics comes into play. A striking observation is the relevance of
the dimensional estimates that are given by Kolmogorov 1941 theory,
even when the fluid flow Reynolds number is so low that no scaling
range can be observed for Eulerian velocity statistics. Understanding
the reasons of such a behaviour requires to investigate in a more
systematic manner the dynamics of particles with inertial-range sizes
in fully developed turbulent flow. Applying the pseudo-penalisation
method to larger particles and higher fluid flow Reynolds numbers is
the subject of on-going work that is mainly focusing on describing the
flow modification induced by the presence of the spherical particle.

\smallskip
\noindent {\sc Acknowledgments.} This study benefited from fruitful
discussions with M.~Bourgoin, E.~Calzavarini, R.~Pasquetti, and
Y.\ Ponty.  This research was supported by the Agence Nationale de la
Recherche under grant No.\ BLAN07-1\_192604.  During his stay in Nice,
H.\ Homann benefitted from a grant of the DAAD. Access to the IBM
BlueGene/P computer JUGENE at the FZ J\"ulich was made available
through project HBO22. Part of the computations were performed on the
``m\'{e}socentre de calcul SIGAMM'' and using HPC resources from
GENCI-IDRIS (Grant 2009-i2009026174).

\bibliographystyle{jfm}
\bibliography{bib}

\end{document}